\documentclass{article}
\usepackage{amsmath,amsthm}
\usepackage{graphicx,epstopdf,epsfig,multirow,epic,bm}
\usepackage{subfigure}
\usepackage{amsfonts}

\normalsize \rm
\begin{document}

\begin{center}
{\Large  \textbf { Counterexample: scale-free networked graphs with invariable diameter and density feature }}\\[12pt]
{\large Fei Ma$^{a,}$\footnote{~The author's E-mail: mafei123987@163.com. },\quad Xiaomin Wang$^{a,}$\footnote{~The author's E-mail: wmxwm0616@163.com.} \quad  and  \quad  Ping Wang$^{b,c,d,}$\footnote{~The corresponding author's E-mail: pwang@pku.edu.cn.} }\\[6pt]
{\footnotesize $^{a}$ School of Electronics Engineering and Computer Science, Peking University, Beijing 100871, China\\
$^{b}$ National Engineering Research Center for Software Engineering, Peking University, Beijing, China\\
$^{c}$ School of Software and Microelectronics, Peking University, Beijing  102600, China\\
$^{d}$ Key Laboratory of High Confidence Software Technologies (PKU), Ministry of Education, Beijing, China}\\[12pt]
\end{center}

\begin{quote}
\textbf{Abstract:} Here, we propose a class of scale-free networks $G(t;m)$ with some intriguing properties, which can not be simultaneously held by all the theoretical models with power-law degree distribution in the existing literature, including (i) average degrees $\langle k\rangle$ of all the generated networks are no longer a constant in the limit of large graph size, implying that they are not sparse but dense, (ii) power-law parameters $\gamma$ of these networks are precisely calculated equal to $2$, as well (iii) their diameters $D$ are all an invariant in the growth process of models. While our models have deterministic structure with clustering coefficients equivalent to zero, we might be able to obtain various candidates with nonzero clustering coefficient based on original networks using some reasonable approaches, for instance, randomly adding some new edges under the premise of keeping the three important properties above unchanged. In addition, we study trapping problem on networks $G(t;m)$ and then obtain closed-form solution to mean hitting time $\langle \mathcal{H}\rangle_{t}$. As opposed to other previous models, our results show an unexpected phenomenon that the analytic value for $\langle \mathcal{H}\rangle_{t}$ is approximately close to the logarithm of vertex number of networks $G(t;m)$. From the theoretical point of view, these networked models considered here can be thought of as counterexamples for most of the published models obeying power-law distribution in current study.


\end{quote}

\vskip 1cm

\section{INTRODUCTION}

Complex systems, such as, friendship networks, metabolic networks, protein-protein interaction networks, predator-prey networks, can be naturally interpreted as complex network, a newborn yet useful tool that has been widely adopted in a large variety of disciplines, particularly, in statistic physics and computer science \cite{Newman-2018}. So, in the past two decades, complex networks have attracted considerable attention and helped us to understand some topological properties and structural dynamics on the complex systems mentioned above. Two significant findings of which are the small-word property \cite{Watts-1998} and scale-free feature \cite{Albert-1999-1}.

There are in general two directions in current complex network study. The one is to generate complex networked models, also called synthetic networks, in order to mimic some characters prevalent in real-world networks, such as power-law degree distribution, small-world phenomena, hierarchical structure \cite{Watts-1998}-\cite{Zamani-2018}. The other aims at determining the influence from topological structure of networks on dynamics taking place on networks themselves, for instance, the mean hitting time for trapping problem, synchronization in networks, epidemic spread \cite{Coghi-2019}-\cite{Satorras-2018}. In this paper, we not only propose a family of networked models $G(t;m)$ and discuss some commonly used topological measures for understanding models $G(t;m)$ in much detail, but also consider the trapping problem on the proposed models $G(t;m)$ and final derive the closed-form solution to mean hitting time.

The main concern in the previous research of theoretical networked models is to focus on constructing models which have scale-free and small-world characters as described above. However, an overwhelming number of models are sparse, implying that the average degree of networks will tend to a constant in the limit of large graph size. This is because a great deal of real-life networks are found to display sparsity feature. By contrast, current studies in some areas turn out the existence of dense networks \cite{Blagus-2012}. To describe these such networks, some available networked models have been proposed \cite{Lambiotte-2016} and analytically investigated in some principled manners, including mean-field theory and master equation. Yet, most of them are stochastic. To our knowledge, almost no deterministic models with both density structure and scale-free feature are built in the past. Although there exist some disadvantages inherited by the latter in comparison with stochastic models, the deterministic structure of model allows us to precisely derive the solutions to some quantities of great interest, such as clustering coefficient, degree distribution, average path length. To some extent, determining some invariants on networked models with deterministic structure has the theoretical flavor. Motivated by this, we present a class of novel networked models with hierarchical structure that are precisely proved to not only show scale-free and small-world characters but also be dense. Throughout this paper, all graphs (models) addressed are simple and the terms graph and network are used indistinctly.

The rest of this paper can be organized by the following several sections. Section II aims at introducing networked models and discussing some widely studied structural parameters, including average degree $\langle k\rangle$, diameter $D$. Among of them, while our models are analytically proved to show scale-free feature, the power-law exponents $\gamma$ is equal to an unexpected constant $2$. This suggests that the density feature can be found on our models. Surprisingly, all the networked models have an identical diameter that is always invariable in the growth process ($t\geq1$). These properties above are rarely reported in current complex network studies. In addition, we take some effective measures, for instance, adding edges with some probability $p$, to switch the original graphs with deterministic structure into stochastic ones. Last but not least, we take account into a type of random walks, called trapping problem, and then analytically determine the mean hitting time $\langle \mathcal{H}\rangle_{t}$. The result shows an interesting phenomenon that the closed-form solution to quantity $\langle \mathcal{H}\rangle_{t}$ is asymptotically close to the logarithm of vertex number of our models under consideration, which is not covered by almost all theoretical models characterizing complex networks with scale-free feature. Final, we close this paper with a concise conclusion and some future directions in Section III.

\section{Networked graphs }

The goal of this section is to build up our networked models, denoted by graphs $G(t;m)$, with hierarchical structure and study some topological structural parameters on these proposed models both analytically and experimentally. In addition, we also consider the trapping problem on the proposed models $G(t;m)$ where the trap is allocated on the largest degree vertex.

\subsection{Construction}

\begin{figure*}
\centering
\includegraphics[height=4cm]{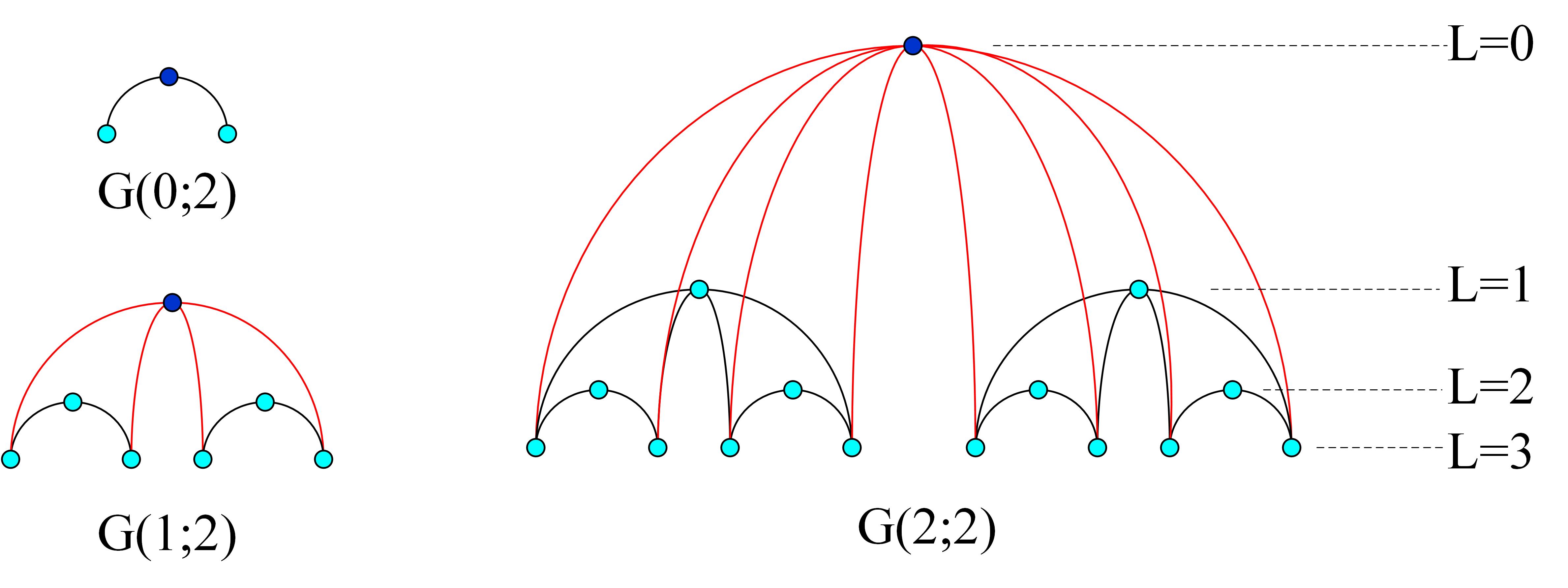}
\caption{\label{fig:wide} The diagram of first three examples of graph $G(t;2)$.}
\end{figure*}

Here we will introduce the graphs $G(t;m)$ ($m\geq2$). First, the seed, denoted by $G(0;m)$, is a star with $m$ leaves as shown in the top-left panel of Fig.1. The next graph $G(1;m)$ can be obtained from $G(0;m)$ in the following manner, (\textbf{1}) generating $m$ duplications of seed $G(0;m)$ labelled as $G^{i}(0;m)$, (\textbf{2}) taking an active vertex, (\textbf{3}) connecting that active vertex to each leaf in star $G^{i}(0;m)$. Obviously, the resulting graph $G(1;m)$ has hierarchical structure as plotted in the bottom-left of Fig.1. For convenience, we divide all vertices of $G(1;m)$ into three classes, i.e., that active vertex allocated at the level $0$, denoted by $L=0$, all the central vertices of stars $G^{i}(0;m)$ at the level $L=1$ and the remaining vertices of graph $G(1;m)$ at the level $L=2$. Henceforth, for time step $t\geq2$, the young graph $G(t;m)$ can be built based on $m$ duplications of the preceding graph $G(t-1;m)$ by connecting an active vertex to all the vertices at the level $L=t$ of graphs $G^{i}(t-1;m)$. As an illustrative example, the graph $G(2;2)$ is shown in the rightmost panel of Fig.1.

In view of the growth manner of graph $G(t;m)$, it is not hard to obtain a couple of equations satisfied by vertex number $|V(t;m)|$ and edge number $|E(t;m)|$ as follows

\begin{equation}\label{eqa:MF-2-1}
\left\{\begin{aligned}&|V(t;m)|=m|V(t-1;m)|+1\\
&|E(t;m)|=m|E(t-1;m)|+m^{t+1}
\end{aligned}\right..
\end{equation}

With the initial conditions $|V(0;m)|=m+1$ and $|E(0;m)|=m$, we can solve for $|V(t;m)|$ and $|E(t;m)|$ from Eq.(\ref{eqa:MF-2-1}) to obtain

\begin{equation}\label{eqa:MF-2-2}
|V(t;m)|=\frac{m^{t+2}-1}{m-1},\quad |E(t;m)|=(t+1)m^{t+1}.
\end{equation}

So far, we already accomplish the construction of our networked models exhibiting hierarchy phenomena. Note that the related work to construction of hierarchical networks has been widely reported in the literature, such as Refs.\cite{Albert-2001} and \cite{Ravasz-2003}. For instance, Ravasz \emph{et al} have generated a class of hierarchical networked models in \cite{Ravasz-2003} and then discussed some structural parameters on them. Whereas, it is worthy noticing that the generative method adopted in this paper is slightly different from that used in \cite{Ravasz-2003}. While the both are manipulated in an iterative manner, our method is based on an idea that the next networked model $G(t;m)$ is created by connecting a portion of vertices of each duplication $G^{i}(t-1;m)$ of the proceeding model $G(t-1;m)$ to a new external vertex instead of a designated vertex in model $G(t-1;m)$. In addition, as will become clear in the coming sections, some properties of our models are not found on those models in \cite{Albert-2001} and \cite{Ravasz-2003}.

The next tasks are to in-detail discuss several common topological properties on graphs $G(t;m)$, for instance, average degree $\langle k\rangle$ and clustering coefficient $\langle C\rangle$, as tried in the existing literature \cite{Ma-2019-1},\cite{Zou-2019}.

\subsection{Structural properties}

As described above, the deterministic nature of the proposed graphs allows us to evaluate some well-studied topological parameters associated with the underlying structure. This section aims to calculate the closed-form solutions for several structural indices, including average degree $\langle k\rangle$.

\subsubsection{Average degree }

As the simplest yet most important structural parameter, average degree $\langle k\rangle$, defined as the ratio of $2$ times edge number and vertex number, can be adopted to determine whether a given network is sparse or not. In general, almost all published networked models with scale-free feature (discussed later), both stochastic and deterministic, are by definition sparse, that is, the value for $\langle k\rangle$ being finite in the large graph size limit. By contrast, our networked models $G(t;m)$ turn out to be of density due to

\begin{equation}\label{eqa:MF-3-1-0}
\langle k(t;m)\rangle=\frac{2|E(t;m)|}{|V(t;m)|}\sim 2t=O(\ln|V(t;m)|).
\end{equation}

From Eq.(\ref{eqa:MF-3-1-0}), it is clear to see that average degree $\langle k(t;m)\rangle$ is linearly correlated with time step $t$ and no longer a constant compared with those models in \cite{Albert-2001}-\cite{Ma-2019-1}, such as, Apollonian networks \cite{Zhang-2019}. Meantime, as recently reported in \cite{Blagus-2012}, many real-world network examples have been proven to have no sparsity topological structure. Therefore, Our networked models $G(t;m)$ may be able to be selected as potential models to unveil some unseen properties behind those dense networks in real life.

\subsubsection{Degree distribution }

In the past years, there are two significant findings in complex network study. One of which is scale-free feature due to Barabasi and Albert \cite{Albert-1999-1} using statistical method for depicting vertex degree distribution of many real-world networks. Such types of networks show a fact that a small fraction of vertices possess a great number of connections and however the rest of vertices have a small number of connections. Since then, given a complex networked model, one always estimate using degree distribution whether or not it is scale-free. Taking into account deterministic structure of graphs $G(t;m)$, we make use of the cumulative degree distribution in discrete form

\begin{equation}\label{eqa:MF-3-2-0}
P_{cum}(k\geq k_{i})=\frac{\sum_{k\geq k_{i}}N_{k}}{|V|},
\end{equation}
where symbol $N_{k}$ represents the total number of vertices with degree exactly equal to $k$ in graph $G(V,E)$. As said in Eq.(\ref{eqa:MF-3-2-0}), we need to classify all vertices of graphs $G(t;m)$ according to vertex degree in order to determine whether the presented models show scale-free feature.

For a graph $G(t;m)$ with $t+1$ levels, it is evident to see that the greatest degree vertex is that active vertex added at time step $t$ and has degree $m^{t+1}$, the second greatest degree vertices are all at the level $L=1$, and so on. With such a classification, there is in fact an unexpected case where all vertices at level $L=t+1$ have degree $t+1$. This degree value must be in some range between $m^{t_{i}}$ and $m^{t_{i+1}}$. Therefore, we will adjust the initial ranks of vertices with respect to vertex degree alone for simplicity. The end list is as follows

\begin{center}
\begin{tabular}{c|c|c|c|c|c|c|c|cccc}
  \hline
  $L$ & $0$ & $1$ & $...$ & $t_{i}-1$ & $...$ & $t-1$ & $t$ & $t+1$ \\
   \hline
  $k_{L,t;m}$ & $m^{t+1}$ & $m^{t}$ & $...$ & $m^{t_{i}}$ & $...$ & $m^{2}$ & $m$ & $t+1$ \\
   \hline
 $N_{L,t;m}$ & $1$ & $m$ & $...$ & $m^{t-t_{i}+1}$ & $...$ & $m^{t-1}$ & $m^{t}$& $m^{t+1}$ \\
  \hline
\end{tabular}
\end{center}
In practice, the above list is usually called degree sequence of graph $G(t;m)$ in the jargon of graph theory. Based on such a list and Eq.(\ref{eqa:MF-3-2-0}), we can calculate the degree distribution of networked model $G(t;m)$ in the following

\begin{equation}\label{eqa:MF-3-2-1}
P_{cum}(k\geq k_{i})=\left\{
\begin{array}{ll}
\quad k_{i}^{-\gamma_{\alpha}}\; , \qquad & \quad k_{i}> t+1\\
k_{i}^{-\gamma_{\alpha}}+\frac{1}{2}\; , & \quad k_{i}\leq t+1 \\
\end{array}
\right.
\end{equation}
where power-law parameter $\gamma_{\alpha}=1$. Performing the derivative of both sides in Eq.(\ref{eqa:MF-3-2-1}) with respect to $k$ yields
\begin{equation}\label{eqa:MF-3-2-2}
P(k)\sim k^{-\gamma}, \quad \gamma=\gamma_{\alpha}+1=2.
\end{equation}
This disapproves a statement in \cite{Charo-2011} that no network with an unbounded
power-law degree distribution with $0\leq\gamma\leq2$ can exist in the limit of large graph size.

Technically, it is an obvious corollary from Eq.(\ref{eqa:MF-3-2-2}) that our networked models $G(t;m)$ are all dense as in Eq.(\ref{eqa:MF-3-1-0}) because of the feature of Riemann $\zeta(1)$ function

\begin{equation}\label{eqa:MF-3-2-3}
\zeta(1)=\sum_{i=1}i^{-1}.
\end{equation}
In the limit of large $i$, the right-hand side of Eq.(\ref{eqa:MF-3-2-3}) is divergence. Thus, average degree $\langle k(t;m)\rangle$ of networked models $G(t;m)$ must be infinite due to $\langle k(t;m)\rangle=\int_{k_{min}}^{k_{max}}kP(k)dk$.

As showed in \cite{Caldarelli-2004}, Caldarelli \emph{et al} have pointed out the widespread occurrence of the inverse square distribution in social sciences and taxonomy and also provided some detailed discussions about mechanisms causing such a phenomenon. Meanwhile, they showed that the treelike classification method is competent to lead to this behavior mentioned above. At the same time, several stochastic scale-free graphs with degree parameter $\gamma=2$ have been constructed in \cite{Courtney-2018}.  Nonetheless, there are no deterministic models following power-law degree distribution with $\gamma=2$ found in the current theoretical models study. Roughly speaking, our models, graphs $G(t;m)$, can be regarded as the first attempt in context of constructing deterministic models. In addition, graphs $G(t;m)$ have many other interesting topological structural properties as we will show shortly. Some of which are not displayed by those stochastic models. In particular, the finding of ultrasmall diameter is a surprising result in this sense.

\subsubsection{Diameter }

Following the previous subsection, the other intriguing finding in the complex network studies is small-world property attributed to Watts and Strogatz \cite{Watts-1998} by empirically capturing the diameter of some real-world and synthetic network models. Mathematically, diameter of a graph, denoted by $D$, is the maximum over distances of all possible vertex pairs. For a pair of vertices $u$ and $v$, distance between them, denoted by $d_{uv}$, is the edge number of any shorted path joining vertex $u$ and $v$. Most generally, diameter can be viewed as a coarse-granularity index for measuring the information delay on a network in question.

With the help of concrete construction of networked models $G(t;m)$, it is easy to find the diameters $D(t;m)$ to obey

\begin{equation}\label{eqa:MF-3-3-0}
D(t;m)=4=O(1).
\end{equation}
This is because (\textbf{1}) all vertices at level $L=t+1$ are connected to that active vertex at the highest level, namely, $L=0$; (\textbf{2}) each vertex at the intermediate levels, $L=1,...,L=t$, always connects to a vertex at the most bottom level $L=t+1$. In fact, diameters $D(t;m)$ are equal to the distance between that two vertices that are both at the intermediate levels and in different branches of graphs $G(t;m)$. Taking into account a trivial character in a connected graph $G(V,E)$ that average path length, defined as $\langle d\rangle=\sum_{u,v\in E, u\neq v}d_{uv}/\frac{1}{2}[|V|(|V|-1)]$, is no larger than the diameter, we omit the analytical solutions for average path length of networked models $G(t;m)$. Using the exact value of diameter in Eq.(\ref{eqa:MF-3-3-0}), one can be convinced that our graphs have captured an ingredient of small-world property. The other will be discussed in the subsequent section.

It is worthy noting that in \cite{Cohen-2003}, the authors have reported some results about diameter of scale-free graphs $G(V,E)$, for instance, those with $2<\gamma<3$ having a much smaller diameter in the limit of large graph size, which behaves as $D\sim \ln\ln|V|$. However, there are few discussions about diameter of the scale-free graphs with $\gamma=2$ in published papers. Perhaps one of important reasons for this is that the previous researches focus mainly on sparse models with scale-free feature. Here, our networked models $G(t;m)$ are proved to have an invariable diameter in the evolution process. As a result, graphs $G(t;m)$ can serve as stronger evidences for illustrating that our ability to understand the fundamental structural properties of graphs including all scale-free ones is always limited to some specific models and hence some early demonstrations may be not complete.

Compared to numerous pre-existing networked models with scale-free feature, such as those in \cite{Cohen-2003}, the power-law parameter $\gamma$ of our graphs $G(t;m)$ is not in the range $2<\gamma<3$. Hence, one has mostly likely to conjecture whether the density feature of networked models $G(t;m)$ leads to such an ultrasmall diameter directly. In principle, the dense graphs should show smaller diameters in comparison with those sparse ones. In practice, we would like to note that density feature planted on graphs $G(t;m)$ indeed shrinks the distance between any pair of vertices and thus has vastly effect on emergence of smaller diameter, but it is not a sufficient condition. As described in our recent work \cite{Ma-2019}, the diameter of dense graphs $G(V,E)$ obeying power-law distribution may also be quite large compared with the widely-used value $\ln|V|$. Again, this strongly means that there are a great number of structural properties of scale-free graphs incompletely uncovered until now and so more efforts should be paid to better understand this kinds of fascinating graphs in the future.

\subsubsection{Clustering coefficient }

The other ingredient of small-world property is clustering coefficient that plays an important role in evaluating the level of clusters in a networked model under consideration. For instance, it in essence describes a phenomenon that in a friendship network two arbitrary friends of one person will have a higher likely to be friends with each other, usually called the triadic in social analysis. For the theoretical point of view, such a connection trend among neighbors of one vertex $v$ with degree $k$ can be abstractly depicted in the following form $c_{v}=n_{v}/\frac{1}{2}[k(k-1)]$ where $n_{v}$ represents the actually existing edges between neighbors of vertex $v$. For the whole graph $G(V,E)$, the clustering coefficient $\langle C\rangle$ can be defined as the averaged value over clustering coefficients of all vertices, as follows

\begin{equation}\label{eqa:MF-3-4-0}
\langle C\rangle=\frac{\sum_{v\in V}n_{v}}{|V|}.
\end{equation}

\begin{figure}
\centering
\includegraphics[height=6cm]{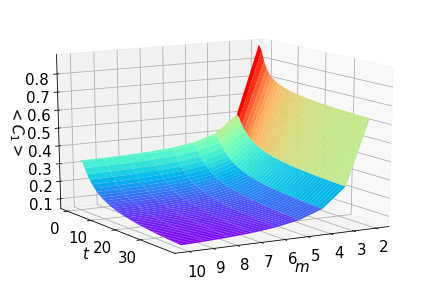}
\caption{\label{fig:wide} The diagram of clustering coefficients $\langle C_{1}\rangle$ of graphs $G_{1}(t;m)$.}
\end{figure}

\begin{figure*}
\centering
\subfigure[]{
\begin{minipage}[t]{0.25\linewidth}
\centering
\includegraphics[width=4.5cm]{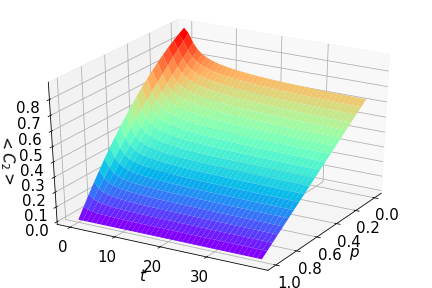}
\end{minipage}%
}%
\subfigure[]{
\begin{minipage}[t]{0.25\linewidth}
\centering
\includegraphics[width=4.5cm]{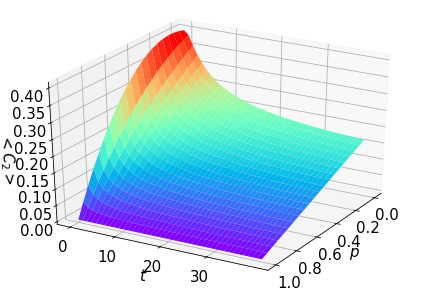}
\end{minipage}%
}%
\subfigure[]{
\begin{minipage}[t]{0.25\linewidth}
\centering
\includegraphics[width=4.5cm]{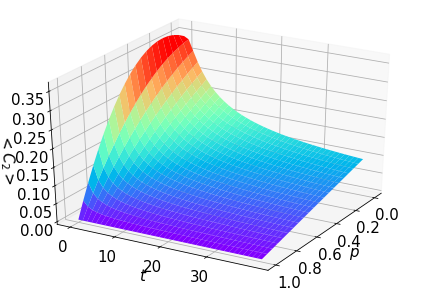}
\end{minipage}
}%
\subfigure[]{
\begin{minipage}[t]{0.25\linewidth}
\centering
\includegraphics[width=4.5cm]{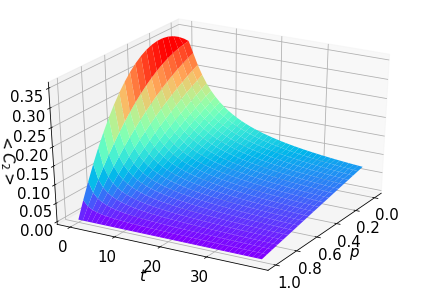}
\end{minipage}
}%
\caption{\label{fig:wide} The diagram of clustering coefficients $\langle C_{2}\rangle$ of graphs $G_{2}(t;m)$ where parameters $m$ are equal to $2,4,6$ and $8$ from left to right, respectively. }
\end{figure*}

By definition, we can without difficulty obtain that the clustering coefficients $\langle C(t;m)\rangle$ of networked models $G(t;m)$ are equivalent to $0$ because there are no triangles contained in the evolution of graphs $G(t;m)$. This suggests that small-world property can not be found in graphs $G(t;m)$. Nevertheless, there are still plenty of other potential characters behind networked models $G(t;m)$. To be more concrete, all the generated graphs $G(t;m)$ are planar, meaning that for an arbitrary parameter $m$, graph $G(t;m)$ can be embedded in the plane so that its edges intersect only at their ends \cite{Bondy-2008}.

To make our models $G(t;m)$ small-world, i.e., achieving the transformation from zero clustering coefficient to nonzero, we have to take some effective measures. There are in fact a great deal of methods for addressing problems of this kind in present research. The simplest one is to add new edges between unconnected vertex pairs for generating enough triangles. Here, our goal is not only to obtain nonzero clustering coefficient but also to remain the properties discussed above unchanged. To this end, we make use of a simple replacement of the seed from star $G(0;m)$ to wheel $G_{1}(0;m)$ \cite{D-2019}. Besides that, all the growth mechanisms keep the same as previously. This leads to a new graph $G_{1}(t;m)$. By definition, it is straightforward to get

\begin{equation}\label{eqa:MF-3-4-1}
\langle C_{1}\rangle=\frac{\sum_{i=0}^{t}\frac{2m^{t-i}}{m^{i+1}-1}+\frac{2(t+1)m^{t+1}}{(t+3)(t+2)}}{|V_{1}(t;m)|},
\end{equation}
where the vertex number $|V_{1}(t;m)|$ of graph $G_{1}(t;m)$ is equal to $|V(t;m)|$. In the large graph size limit, the clustering coefficient $\langle C_{1}\rangle$ of Eq.(\ref{eqa:MF-3-4-1}) tends to a nonzero constant for small parameters $m$ as shown in Fig.2, suggesting that deterministic graph $G_{1}(t;m)$ will have scale-free feature and small-world property simultaneously.

To make further progress, we may delete with probability $p$ each edge between vertices at the level $L=t+1$ in graph $G_{1}(t;m)$, leading to another graph $G_{2}(t;m)$. The introduction of randomly deleting edges will switch deterministic graphs into the opposite case, namely, stochastic ones. As before, the clustering coefficient $\langle C_{2}\rangle$ of stochastic graphs $G_{2}(t;m)$ can be calculated as follows

\begin{equation}\label{eqa:MF-3-4-2}
\langle C_{2}\rangle=\frac{(1-p)\sum_{i=0}^{t}\frac{2m^{t-i}}{m^{i+1}-1}+\left[(1-p)^{2}\frac{2(t+1)}{(t+3)(t+2)}+\frac{2p(1-p)}{t+2}\right]m^{t+1}}{|V_{2}(t;m)|},
\end{equation}

here $|V_{2}(t;m)|$ is the vertex number of graphs $G_{2}(t;m)$. To determine the tendency of $\langle C_{2}\rangle$ in the limit of large size, we feed graphs $G_{2}(t;m)$ into computer. As plotted in Fig.3, for distinct parameters $m$, the values $\langle C_{2}\rangle$ are different from each other initially but all show similar tendency in the large-$t$ limit.

In a word, the three types of networked models, graphs $G(t;m)$, $G_{1}(t;m)$ and $G_{2}(t;m)$, can be selected as counterexamples for disproving some previous statements about scale-free graphs, such as the scale-free graphs with small-world property may have invariable diameter.
Meantime, the lights shed by them may be helpful to construct new networked models in the future.

\subsubsection{Assortative structure }

Many real-world networks \cite{Newman-2002} have been observed to show assortative mixing on their degrees, that is, a preference for high-degree vertices to attach to other vertices like them, while others show disassortative mixing, i.e., high-degree vertices attach to ones unlike them. Particularly, this is a popular phenomenon in social networks. For instance, it is mostly willing of people to establish friendships with those at the same level as them rather than to get in touch with others. For the purposes of quantifying this feature of networks $G(V,E)$, Newman defined the following measure $r$, usually called assortativity coefficient,

\begin{equation}\label{eqa:MF-3-5-0}
r=\frac{|E|^{-1}\sum\limits_{e_{ij}\in E} k_{i}k_{j}-\left[|E|^{-1}\sum\limits_{e_{ij}\in E} \frac{1}{2}(k_{i}+k_{j})\right]^{2}}{|E|^{-1}\sum\limits_{e_{ij}\in E} \frac{1}{2}(k^{2}_{i}+k^{2}_{j})-\left[|E|^{-1}\sum\limits_{e_{ij}\in E} \frac{1}{2}(k_{i}+k_{j})\right]^{2}},
\end{equation}
in which $k_{i}$ is the degree of vertex $v$ and $e_{ij}$ denotes an edge connecting vertex $i$ to $j$. With such an index, most social networks turn out to have significant assortative mixing, while technological and biological networks seem to be disassortatively constructed.

As explained above, the scalar measure $r$ in fact figures the degree of similarity between two endpoints of any edge on an observed network by means of vertex degree. Empirically, our networked models $G(t;m)$ should be disassorative. To show this, we can write based on Eq.(\ref{eqa:MF-3-5-0})

\begin{figure}
\centering
\includegraphics[height=5cm]{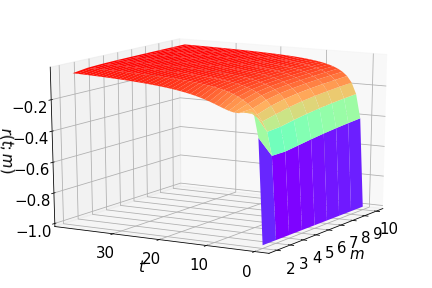}
\caption{\label{fig:wide} The diagram of assortativity coefficient $r(t;m)$ of graphs $G(t;m)$.}
\end{figure}

\begin{equation}\label{eqa:MF-3-5-1}
r(t;m)=\frac{\frac{m^{t+2}-m}{m-1}-\left[\frac{t+1}{2}+\frac{m^{t+2}-m}{(2m-2)(t+1)}\right]^{2}}{\frac{(t+1)^{2}}{2}+\frac{m^{2t+4}-m^{2}}{(2m^{2}-2)(t+1)}-\left[\frac{t+1}{2}+\frac{m^{t+2}-m}{(2m-2)(t+1)}\right]^{2}}.
\end{equation}
As $t\rightarrow\infty$, for a given parameter $m$, $r(t;m)$ tends to zero, seeing Fig.4 for a lot. By analogy with Eq.(\ref{eqa:MF-3-5-1}), we can evaluate assortativity coefficient $r_{2}(t;m)$ of stochastic graphs $G_{2}(t;m)$ by plugging the following equations into Eq.(\ref{eqa:MF-3-5-0})
\begin{subequations}
\label{eq:whole}
\begin{eqnarray}
|E|=|E_{2}(t;m)|=m^{t+1}(t+2-p),\label{subeq:MF-3-5-2}
\end{eqnarray}
\begin{equation}
\begin{aligned}\sum\limits_{e_{ij}\in E} k_{i}k_{j}&=m^{t+1}\sum_{i=0}^{t}[p^{2}(t+1)+2p(1-p)(t+2)]\\
&+m^{t+1}\sum_{i=0}^{t}(1-p)^{2}(t+3)+(1-p)^{4}(t+3)^{2}m^{t+1}\\
&+4p(1-p)^{2}[p(t+2)^{2}+(1-p)(t+2)(t+3)]m^{t+1}
\end{aligned},\label{subeq:MF-3-5-3}
\end{equation}
\begin{equation}
\begin{aligned}\sum\limits_{e_{ij}\in E}(k_{i}+k_{j})&=m^{t+1}(t+1)+\sum_{i=0}^{t}(1-p)^{2}(t+3)m^{i} \\
&+\sum_{i=0}^{t}m^{i}[p^{2}(t+1)+2p(1-p)(t+2)]\\
&+4p(1-p)^{2}[p(2t+4)+(1-p)(2t+5)]m^{t+1}\\
&+(1-p)^{4}(2t+6)m^{t+1}
\end{aligned},\label{subeq:MF-3-5-4}
\end{equation}
\begin{equation}
\begin{aligned}\sum\limits_{e_{ij}\in E}(k_{i}^{2}+k_{j}^{2})&=m^{t+1}\sum_{i=0}^{t}m^{i+1}+\sum_{i=0}^{t}(1-p)^{2}(t+3)^{2}m^{i} \\
&+\sum_{i=0}^{t}m^{i}[p^{2}(t+1)^{2}+2p(1-p)(t+2)^{2}]\\
&+2(1-p)^{2}[4p^{2}(t+2)^{2}+(1-p)^{2}(t+3)^{2}]m^{t+1}\\
&+4p(1-p)^{3}[(t+2)^{2}+(t+3)^{2}]m^{t+1}
\end{aligned}.\label{subeq:MF-3-5-5}
\end{equation}
\end{subequations}

In order to evaluate whether the random deletion of edges has influence on assortativity coefficient $r_{2}(t;m)$, we conduct extensive simulations in terms of Eq.(\ref{eqa:MF-3-5-0}) and Eqs.(\ref{subeq:MF-3-5-2})-(\ref{subeq:MF-3-5-5}) and experimental results are shown in Fig.5.

Interestingly, from the panels in Fig.5, it can be easy to see a phenomenon that all theoretical values for assortativity coefficients $r_{2}(t;m)$ are bounded from above the critical condition $0$ while approaching to zero in the large graph size limit. This is sharply different from those previously reported results in \cite{Newman-2002} where all assortativity coefficients associated with most of studied networked models are negative while also tending to zero.

\begin{figure*}
\centering
\subfigure[]{
\begin{minipage}[t]{0.25\linewidth}
\centering
\includegraphics[width=4.8cm]{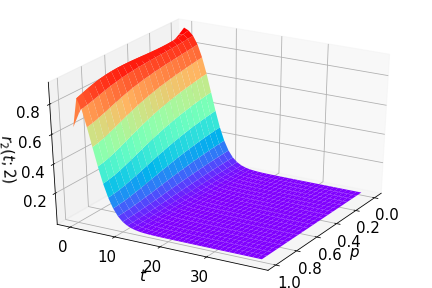}
\end{minipage}%
}%
\subfigure[]{
\begin{minipage}[t]{0.25\linewidth}
\centering
\includegraphics[width=4.8cm]{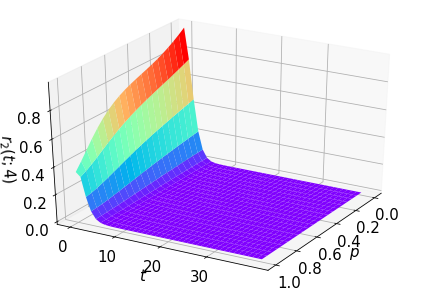}
\end{minipage}%
}%
\subfigure[]{
\begin{minipage}[t]{0.25\linewidth}
\centering
\includegraphics[width=4.8cm]{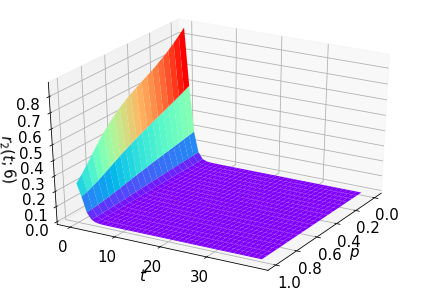}
\end{minipage}
}%
\subfigure[]{
\begin{minipage}[t]{0.25\linewidth}
\centering
\includegraphics[width=4.8cm]{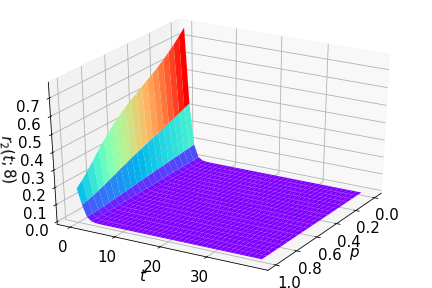}
\end{minipage}
}%
\caption{\label{fig:wide} The diagram of assortativity coefficients $r_{2}(t;m)$ of graphs $G_{2}(t;m)$ where parameters $m$ are equal to $2,4,6$ and $8$ from left to right, respectively. }
\end{figure*}

\subsubsection{Mean hitting time}

In this section, we formulate the trapping problem on our graphs $G(t;m)$. In practice, this is a simple unbiased Markovian random walk with a tap, say a perfect absorber, allocated on a designated vertex on a graph in question. As shown above, that active vertex at the level $L=0$ of graphs $G(t;m)$ has the largest degree and hence is called the hub vertex, denoted by $h_{t}$. In order to further probe its importance under trapping problem, we put an absorber on the hub vertex. And then, a particle located on vertex $v$ but for the hub will hop to one of its neighbor $N_{v}$ with the transition probability $1/d_{v}(t;m)$ before arriving at that absorber where $d_{v}(t;m)$ is the degree of vertex $v$ in graphs $G(t;m)$.

Consider that a particle starts from vertex $v$ at initial time, the jumping probability $P_{vu}$ of starting out from $v$ to $u$ satisfies the following master equation

\begin{equation}\label{eqa:MF-3-6-0}
P_{vu}(l+1)=\sum_{i\in V(t;m)}\frac{a_{iu}}{d_{i}(t;m)}P_{vi}(l),
\end{equation}
where $a_{iu}$ is the element of adjacency matrix of graph $G(t;m)$, $a_{iu}=1$ if this pair of vertices $i$ and $u$ are connected by an edge and $a_{iu}=0$ otherwise.

According to the rule above, we are particularly interested in the quantity, called hitting time $\mathcal{H}$, for measuring the expected time for a particle, which starts from an arbitrary vertex, to first visit at the trap in the trapping problem. For graph $G(t;m)$ as a whole, we denote the hitting time for a particle placed on vertex $v$ by $\mathcal{H}_{v}$ and then let $P(\mathcal{H}_{v}=l)$ be the probability for that particle to first hit the trap, i.e., hub vertex $h_{t}$, after $l$ steps. By analogy with Eq.(\ref{eqa:MF-3-6-0}), we can obtain

\begin{equation}\label{eqa:MF-3-6-1}
P(\mathcal{H}_{v}=l)=\sum_{i\in V(t;m),i\neq h_{t}}\frac{a_{iu}}{d_{i}(t;m)}P(\mathcal{H}_{v}=l-1).
\end{equation}

A commonly used approach for the preceding equation is generating function. Without loss of generality, we may define the corresponding generating function of quantity $P(\mathcal{H}_{v}=l)$ in the following form

\begin{equation}\label{eqa:MF-3-6-2}
\mathcal{P}_{v}(x)=\sum_{t=0}^{\infty}P(\mathcal{H}_{v}=l)x^{l}.
\end{equation}
As we will show later, a trial yet useful fact related to $\mathcal{P}_{v}(x)$, that is, the expected time $\mathcal{H}_{v}$ is exactly equal to the value $\mathcal{P}'_{v}(1)$, helps us to consolidate all the results in the subsequent section.

Before beginning to derive our calculations, for convenience, we need to introduce two notations $P_{t}(s)$ and $Q_{L}(s)$. The former represents the probability for a particle on an arbitrary vertex at the level $L=t+1$ of graphs $G(t;m)$ to first arrive at the hub $h_{t}$ after $s$ steps, and the latter is defined as the probability that a particle originating from an arbitrary vertex $w$ at the level $L$ ($L=1,\dots,t$) hits one at random chosen vertex at the level $L=t+1$ of graphs $G(t;m)$, which connects to vertex $w$, after $s$ jumps. Based on the structure of graphs $G(t;m)$ and the statements above, we can write the following equation

\begin{equation}\label{eqa:MF-3-6-3}
P_{t}(s)=\frac{\delta_{s,1}}{k_{L=t+1}}+\frac{1}{k_{L=t+1}}\sum_{L=1}^{t}\sum_{i=1}^{s-1}Q_{L}(i)P_{t}(s-1-i),
\end{equation}
here $\delta_{s,1}$ is the Kronecker delta function in which $\delta_{s,1}=1$ as $s=1$ and $\delta_{s,1}=0$ otherwise, $k_{L=t+1}$ is the degree of vertex at the level $L=t+1$ and equals $t+1$ as above.

Using the lights shed by Eq.(\ref{eqa:MF-3-6-2}), the generating function $\mathcal{P}_{t}(x)$ corresponding to quantity $P_{t}(s)$ can be expressed as follows

\begin{equation}\label{eqa:MF-3-6-4}
\mathcal{P}_{t}(x)=\frac{x}{k_{L=t+1}}+\frac{tx^{2}}{k_{L=t+1}}\mathcal{P}_{t}(x),
\end{equation}
in which we make use of an evident result $Q_{L}(i)=1$ only for both $L=1,\dots,t$ and $i=1$, as well $Q_{L}(i)=0$ otherwise.

At the same time, we let $\mathcal{H}_{t+1}^{t}$ stand for the hitting time for a particle initially set on any vertex at the level $L=t+1$ which is by definition written

\begin{equation}\label{eqa:MF-3-6-5}
\mathcal{H}_{t+1}^{t}=\left.\dfrac{d}{dx}\mathcal{P}_{t}(x)\right|_{x=1}.
\end{equation}

Taking into consideration Eq.(\ref{eqa:MF-3-6-5}), performing the derivative of both sides of Eq.(\ref{eqa:MF-3-6-4}) produces the exact solution of $\mathcal{H}_{t+1}^{t}$ as follows

\begin{equation}\label{eqa:MF-3-6-6}
\mathcal{H}_{t+1}^{t}=2t+1.
\end{equation}

For each vertex at the level $L=1,\dots,t$, combining the definition of $Q_{L}(s)$ and the hierarchical structure of graphs $G(t;m)$, the hitting time $\mathcal{H}_{L}^{t}$ for a particle originally allocated on any vertex at the level $L$ may be obtained in terms of $\mathcal{H}_{t+1}^{t}$

\begin{equation}\label{eqa:MF-3-6-7}
\mathcal{H}_{L}^{t}=\mathcal{H}_{t+1}^{t}+1.
\end{equation}

By far, the hitting times $\mathcal{H}_{L}^{t}$ for a particle at the level $L=1,\dots,t+1$ are all precisely calculated in a rigorous manner. The next task is to derive the mean hitting time $\langle \mathcal{H}\rangle_{t}$, which characterizes the trapping process on average, in the following fashion

\begin{equation}\label{eqa:MF-3-6-8}
\langle \mathcal{H}\rangle_{t}=\frac{1}{|V(t;m)|-1}\sum_{L=1}^{t+1}\mathcal{H}_{L}^{t}|N_{t}(L)|,
\end{equation}
here we again utilize the hierarchy of graphs $G(t;m)$ and $|N_{t}(L)|$ denotes the total number of vertices at the level $L$ ($L=1,\dots,t+1$). As pointed out before, $|N_{t}(L)|$ is in essence equal to $N_{L,t;m}$.

Substituting Eqs.(\ref{eqa:MF-3-6-6})-(\ref{eqa:MF-3-6-7}) and the value of $|V(t;m)|$ in Eq.(\ref{eqa:MF-2-1}) into Eq.(\ref{eqa:MF-3-6-8}) yields

\begin{equation}\label{eqa:MF-3-6-9}
\langle \mathcal{H}\rangle_{t}=O(2t+\frac{1}{m}),
\end{equation}
where we take useful advantage of some simple arithmetics.

To make process further, we consider the logarithm of vertex number of graphs $G(t;m)$, namely, $\ln|V(t;m)|\sim t\ln m$. It is clear to see that for the whole graphs $G(t;m)$, the mean hitting time $\langle \mathcal{H}\rangle_{t}$ has a close relationship with the vertex number of graphs $G(t;m)$ as shown below

\begin{equation}\label{eqa:MF-3-6-10}
\langle \mathcal{H}\rangle_{t}\sim\ln|V(t;m)|.
\end{equation}

This is completely different from some previous results in the existing literature, such as, the complete graph on $N$ vertices having the mean hitting time exactly equal to $N-1$, which is quite approximately close to its vertex number, the hierarchical models considered in \cite{Zhang-2009} with the mean hitting time also being the same order of magnitude as their own vertex number. Compared to our models, the former, i.e., complete graph, has no both scale-free feature and hierarchical structure while with the smallest diameter. On the other hand, the models in \cite{Zhang-2009} have a larger enough diameter than our graphs while showing scale-free feature and the hierarchy of structure. Following the discussions above, it has been shown that when the trap is allocated on the greatest degree vertex, those hierarchical networks presented in \cite{Albert-2001} and \cite{Ravasz-2003} have mean hitting time approximately close to the power of number of vertices for some exponent $\alpha$ in form, a demonstration that is distinct with the consequence obtained in this paper. On average, this suggests that the proposed networked models $G(t;m)$ outperform those models \cite{Albert-2001} and \cite{Ravasz-2003} with respect to mean hitting time in the trapping problem considered here.

As a consequence, our models can be capable of serving as counterexamples for in-depth understanding many other fundamental properties on theoretical models, in particular, with respect to scale-free graphs. One of the most important reasons is that scale-free feature is ubiquitously observed in a large amount of complex networks, both synthetic and real-world.

\section{Conclusion}

In summary, we present a family of scale-free networked models of significant interest. Based on both theoretical arguments and experimental simulations, we derive some striking results unseen in pre-existing theoretical models. They shows that (1) our graphs $G(t;m)$ follow power-law degree distribution with exponent $2$ and thus are dense, (2) an invariable diameter can be found on our graphs $G(t;m)$ compared to almost all previously proposed scale-free models, (3) using random method leads the end graphs $G_{2}(t;m)$ to display a nonnegative assortativity coefficient, and (4) when the trap is allocated on the hub in graphs $G(t;m)$, the mean trapping time is approximately related to the logarithm of vertex number of graphs. To the best of our knowledge, this work seems the first to probe novel scale-free models particularly because in the last, a significant amount of attention have been paid to discuss sparse graphs with scale-free feature. From the respect of theoretical research, our models can be used as counterexamples to disprove some previous demonstrations corresponding to the scale-free graph family in current study and so enable researchers to well understand the fundamental structure properties planted on scale-free models.

\section*{Acknowledgments}
The research was supported by the National Key Research and Development Plan under grant 2017YFB1200704 and the National Natural Science Foundation of China under grant No.
61662066.


\begin{thebibliography}{9}
\bibitem{D-2019} The wheel graph $G_{1}(0;m)$ can be obtained from star $G(0;m)$ by adding new edges to connect all leaves in a manner that generates a cycle $C_{m}$.
\end{thebibliography}
\end{document}